\documentclass[12pt]{article}

\usepackage{scicite}

\newcommand{\be}{\begin{eqnarray}}
\newcommand{\ee}{\end{eqnarray}}
\newcommand{\ket}[1]{\ensuremath{\left| {#1} \right>}}
\newcommand{\bra}[1]{\ensuremath{\left< {#1} \right|}}

\newcommand{\create}{\ensuremath{\,\hat{a}^{\dagger}}}
\newcommand{\destroy}{\ensuremath{\,\hat{a}}}

\newcommand{\splus}{\ensuremath{\hat{\sigma}_+}\,}

\usepackage{latexsym}
\usepackage{graphics}
\usepackage{multirow}
\usepackage{textcomp}

\usepackage{epstopdf}
\usepackage{graphicx}
\usepackage{caption}
\usepackage{subcaption}
\usepackage{pdflscape}

\newenvironment{sciabstract}{%
\begin{quote} \bf}
{\end{quote}}

\newcounter{lastnote}

\title{Quantum harmonic oscillator state synthesis by reservoir engineering}

\author
{D.~Kienzler$^\ast$, H.-Y.~Lo, B.~Keitch, L.~de~Clercq, F.~Leupold,\\
 F.~Lindenfelser, M.~Marinelli, V.~Negnevitsky, J.~P.~Home$^\ast$\\
\\
\normalsize{Institute for Quantum Electronics, ETH Z\"urich}\\
\normalsize{Otto-Stern-Weg 1, 8093 Z\"urich, Switzerland}\\
\\
\normalsize{$^\ast$To whom correspondence should be addressed;} \\
\normalsize{E-mail: daniel.kienzler@phys.ethz.ch, jhome@phys.ethz.ch}
}

\date{}

\begin{document}




\maketitle


\begin{sciabstract}
The robust generation of quantum states in the presence of decoherence is a primary challenge for explorations of quantum mechanics at larger scales. Using the mechanical motion of a single trapped ion, we utilize reservoir engineering to generate squeezed, coherent and displaced-squeezed states as steady states in the presence of noise. We verify the created state by generating two-state correlated spin-motion Rabi oscillations resulting in high contrast measurements. For both cooling and measurement, we use spin-oscillator couplings that provide transitions between oscillator states in an engineered Fock state basis. Our approach should facilitate studies of entanglement, quantum computation, and open-system quantum simulations in a wide range of physical systems.

\end{sciabstract}

\noindent
Reservoir engineering is a method in which specially designed couplings between a system of interest and a zero temperature environment can be used to generate quantum superposition states of the system as the steady state of the dissipative process, independent of the initial state of the system \cite{93Cirac2,96Poyatos,01Carvalho}.
Theoretical work has shown the potential for using such engineered dissipation for universal quantum computation \cite{08Verstraete}, and in providing new routes to many-body states \cite{08Diehl,08Kraus,11Pastawski}. Experimentally, these techniques have been used to generate entangled superposition states of qubits in atomic ensembles \cite{11Krauter}, trapped ions \cite{11Barreiro,13Lin} and superconducting circuits \cite{13Shankar}. Theoretical proposals for quantum harmonic oscillator state synthesis using reservoir engineering extend from trapped ions \cite{96Poyatos,01Carvalho} to superconducting cavities \cite{11Sarlette,14NavarreteBenlloch} and nano-mechanics \cite{13Kronwald}.

Here, we experimentally demonstrate the generation and stabilization of quantum harmonic oscillator states by reservoir engineering following the original proposal of Cirac \emph{et al.} \cite{93Cirac2}, which we use to generate and stabilize squeezed, displaced-squeezed and coherent states. Making use of engineered spin-motion couplings that are closely related to those used in the reservoir engineering, we introduce measurement techniques that provide simple spin population dynamics, allowing us to directly verify the coherence of the states produced and providing a measure of the fidelity with high signal-to-noise.
$\left|n\right>$
The energy eigenstates of the harmonic oscillator $\left|n\right>$ form an equally spaced ladder connected by the action of the creation and annihilation operators $\hat{a}^{\dagger}$ and $\hat{a}$. Alternative state ladders exist, in which each state is a superposition of energy eigenstates. These can be obtained by applying a unitary transformation, with the resulting states $\left|\hat{U}, n\right> \equiv \hat{U}\left|n\right>$ (Fig. 1A). The transformed state ladders have their own annihilation operators $\hat{K}$, which are related to $\hat{a}$ by  $\hat{K} = \hat{U}\hat{a} \hat{U}^\dagger$ (the same transformation can be performed for the creation operator). State preparation by reservoir engineering involves the choice of a suitable engineered basis for which cooling to the ground state results in the desired quantum state $\left|\hat{U}, 0\right>$. In our experiments, we can cool in this basis by coupling the oscillator to an ancilla spin. We use an engineered spin-motion coupling Hamiltonian
\be
\hat{H}_- = \hbar \Omega\left(\hat{K} \hat{\sigma}_- + \hat{K}^\dagger \hat{\sigma}_-\right) \label{eq:Hc},
\ee
where $\Omega$ is the Rabi frequency and $\hat{\sigma}_- \equiv \left|\uparrow\right>\left<\downarrow\right|,\ \hat{\sigma}_- \equiv \left|\downarrow\right>\left<\uparrow\right|$ are spin flip operators. This Hamiltonian results in transitions between adjacent levels on the transformed state ladder, correlated with spin flips. The Hamiltonian dynamics are reversible, and thus cannot reduce entropy. In order to produce a zero-entropy pure state from a general starting state dissipation is required, which we introduce by optical pumping of the spin. This pumps the oscillator down the engineered state ladder into the desired ground state (Fig. 1B).

We generate Gaussian oscillator states, which are related to the energetic ground state by combinations of displacements and squeezing of the wave packet \cite{BkSchleich}. The unitary transformation is then $\hat{U} = \hat{S}(\xi)\hat{D}(\alpha)$, where $\hat{S}(\xi)$ is the squeezing operator and $\hat{D}(\alpha)$ is the displacement operator \cite{Methods}. The resulting annihilation operator in the engineered basis is $\hat{K} = e^{i\phi}\left(\cosh(r) \hat{a} + e^{i\phi_{\rm s}} \sinh(r) \hat{a}^{\dagger} - \alpha\right)$, where $r = |\xi|$ and $\phi_{\rm s} = \arg(\xi)$. The phase factor $\phi$ plays no role in our experiments and we set it to zero in the rest of the Report. $\hat{K}$ contains terms which are linear in the creation and annihilation operators for the energy eigenstates, meaning that the Hamiltonian $\hat{H}_-$ can be implemented by simultaneously applying a combination of the carrier  ($\hat{H}_{\rm c} = \hbar \Omega_{\rm c} \hat{\sigma}_- + {\rm h.c.}$), red motional sideband ($\hat{H}_{\rm rsb} = \hbar \Omega_{\rm rsb} \hat{\sigma}_- \hat{a} +  {\rm h.c.}$) and blue motional sideband ($\hat{H}_{\rm bsb} = \hbar \Omega_{\rm bsb} \hat{\sigma}_- \hat{a}^{\dagger} + {\rm h.c.}$) transitions. Here $\Omega_{\rm c}, \Omega_{\rm rsb}$ and $\Omega_{\rm bsb}$ are taken to be complex parameters, containing both the coupling strength and the phase. In our experiments these Hamiltonians can be realized simultaneously by applying a multi-frequency laser field, with each frequency component resonant with one of the transitions \cite{Methods}. The squeezing is obtained from the ratio $\Omega_{\rm bsb}/\Omega_{\rm rsb} = e^{i \phi_{\rm s}} \tanh(r)$ and the displacement from the ratio $\Omega_{\rm c}/\Omega_{\rm rsb} = - \alpha/\cosh(r)$.

The experiments work with a single $^{40}\rm{Ca}^{+}$ ion, which oscillates along a chosen direction with a frequency close to $\omega_z/(2 \pi)$ = 1.9~MHz.
At the start of each experimental run, the ion is initialized by cooling all motional modes close to the Doppler limit
using laser light resonant with the dipole transitions at 397 and 866~nm.
All coherent manipulations (including the Hamiltonians used for reservoir engineering) make use of the quadrupole transition at 729~nm, isolating a two-state pseudo\-spin system which we identify as $\left|\downarrow\right> \equiv \left|L = 0, J = 1/2, M_J = +1/2\right>$ and $\left|\uparrow\right> \equiv \left|L = 2, J = 5/2, M_J = 3/2\right>$ \cite{ThRoos}.
The Lamb-Dicke factor for our laser configuration is $\eta = 0.05$, which means that the experiments are well described by the Lamb-Dicke approximation \cite{98Wineland2}.
Optical pumping to $\left|\downarrow\right>$ is implemented using a combination of linearly polarized light fields at 854, 397 and 866~nm \cite{Methods}. The internal electronic state of the ion is read out using state-dependent fluorescence \cite{Methods}.

The reservoir engineering is applied directly after the end of the Doppler cooling. We subsequently probe whether the state of the system has reached the dark state for the Hamiltonian $\hat{H}_-$ by optical pumping into $\left|\downarrow\right>$, followed by a probe pulse using $\hat{H}_-$. Examples of data for a coherent ($\Omega_{\rm bsb}$ = 0) and a squeezed state ($\Omega_{\rm c} = 0$) are shown in Fig. 2. In both cases the state approaches a steady state which approximates a dark state of the pumping Hamiltonian. Experimentally we have implemented two different methods of dissipative pumping. In the first (used for the coherent state data), we repeat a ``cycle'' involving applying $\hat{H}_-$ for a fixed duration followed by repumping of the internal state. The second method (used for the squeezed state data) involves continuous application of both $\hat{H}_-$ and the spin dissipation.
We observe that the motional state pumping is faster in the continuous case. The pulsed method is easier to maintain, as it avoids AC-Stark shifts arising from the repumping laser. It also allows the use of shaped pulses to produce sideband transitions while avoiding undesired off-resonant excitation of the carrier transition \cite{08RoosGate}.

The onset of the dark state indicates that the desired steady state has been reached. We independently verify this state using two methods. The first is a measurement of the populations of the energy eigenstates \cite{96Meekhof}, which we perform by pumping the state to $\left|\downarrow\right>$ after the end of the reservoir engineering and applying the blue sideband Hamiltonian. The population of the state $\left|\downarrow\right>$ as a function of the blue sideband pulse duration $t$ is given by
\begin{equation}
P(\downarrow) = \frac{1}{2}\sum_n p(n) (1 + e^{-\gamma_n t}\cos(\Omega_{n, n + 1}t)) ,
\label{eq:Pflop}
\end{equation}
where $p(n)$ is the probability for finding the oscillator in the $n$th energy eigenstate and $\Omega_{n, n + 1}$ is the Rabi frequency for the transition between the $\left|\downarrow\right>\left|n\right>$ and $\left|\uparrow\right>\left|n + 1\right>$ states, which in the Lamb-Dicke regime scales as $\left<n\right|\hat{a}\left|n+1\right> = \sqrt{n+1}$ \cite{98Wineland2, 96Meekhof}. The phenomenological decay parameter $\gamma_n$ accounts for decoherence and fluctuations in the applied laser intensities \cite{Methods,96Meekhof}. By fitting a form similar to equation \ref{eq:Pflop} to each set of data we obtain the probability distribution $p(n)$ \cite{Methods}. We then fit $p(n)$ using the theoretical form of the probability distributions for coherent, squeezed and displaced-squeezed states \cite{76Yuen}. The data, deduced populations and fits are shown in Fig. 3. The fit for the coherent state yields a coherent state parameter $\left|\alpha\right| = 2.00\pm0.01$ (error bars here and elsewhere are given as SEM). For the squeezed state we obtain a squeezing amplitude $r = 1.45\pm0.03$, which for a pure state would correspond to a $12.6\pm0.3$~dB reduction in the squeezed quadrature variance. For the displaced-squeezed state we obtain fitted parameters of $r = 0.63\pm0.06$, $\left|\alpha\right| = 2.2\pm0.2$ and $\arg{(\alpha)} - \phi_s/2 = 0.42\pm0.06$~rad.

The blue sideband method does not allow us to measure the fidelity of the states, since it does not verify the phase coherence of the superposition. It is also difficult to obtain good population estimates for states with a large spread in their Fock state occupancies \cite{Methods}.
In order to overcome these limitations, we introduce a diagnosis method which provides a Rabi frequency decomposition in the transformed state ladder which includes the desired state as the ground state. Instead of driving only the blue sideband, we use the Hamiltonian
\be
\hat{H}_{+} = \hbar \Omega \left( \hat{K}^\dagger \hat{\sigma}_- + \hat{K} \hat{\sigma}_- \right)
\ee
in which the motional state operators are conjugated with respect to $\hat{H}_-$ (Fig. 1c). This results in Rabi oscillations between the states $\left|\downarrow\right>\left|\hat{U}, n\right>$ and $\left|\uparrow\right>\left|\hat{U}, n+1\right>$. Since the internal states involved span a two-dimensional Hilbert space, the motional state evolution is also contracted onto two adjacent states of the engineered basis. For an arbitrary initial state, the internal state populations evolve according to equation \ref{eq:Pflop}, with the corresponding $p(n)$ being the probability to find the ion in the $n$th element of the engineered basis prior to the application of $\hat{H}_+$ (we denote this as $p_U(n)$ in the figure to avoid confusion). Data sets from this type of measurement are shown for the coherent state and for the squeezed state in Fig. 4 for the same settings as used in figures 2 and 3. In order to work in the same basis as the state engineering, we again drive combinations of the carrier and red and blue motional sidebands, but now with the ratios of Rabi frequencies calibrated according to $\Omega_{\rm c}/\Omega_{\rm bsb} = -\alpha^*/\cosh(r)$ and $\Omega_{\rm rsb}/\Omega_{\rm bsb} = e^{-i \phi_{\rm s}} \tanh(r) $ with $\xi$ and $\alpha$ corresponding to the values used for the reservoir engineering \cite{Methods}.
We fit both experimental data sets with a form similar to equation \ref{eq:Pflop}, obtaining the probability to be found in the ground state of $0.90\pm0.02$ and $0.88\pm0.02$ for the coherent and squeezed states respectively. We take these to be lower bounds on the fidelity with which these states were prepared, since these numbers include errors in the analysis pulse in addition to state-preparation errors \cite{Methods}.
The $\hat{H}_+$ Rabi oscillations observed in our experiments involve transitions which when viewed in the energy eigenstate basis couple Hilbert spaces which are of appreciable size. In order to account for 88\% of the populations in oscillations between $\left|\hat{S}(\xi),0\right>$ and $\left|\hat{S}(\xi),1\right>$ for $r = 1.45$ we must include energy eigenstates up to $n = 26$. By our choice of basis, we reduce the relevant dynamics to a two-state system, greatly simplifying the resulting evolution of the spin populations and thus providing a high signal to noise.
The high fidelity with which the squeezed state is produced is a result of the robust nature of the reservoir engineering, which is insensitive to laser intensity and frequency fluctuations that are common to all frequency components of the engineered Hamiltonian. To generate the same state produced above using standard methods involving unitary evolution starting from the ground state would require simultaneously driving both second motional sidebands \cite{Methods}. We would not expect a high fidelity since these have Rabi frequencies comparable to our transition linewidth, which is broadened by magnetic field fluctuations.

This toolbox for generating, protecting and measuring quantum harmonic oscillator states is transferrable to any physical system in which the relevant couplings can be engineered, facilitating quantum computation with continuous variables \cite{00Gottesman2}. Examples in which reservoir engineering have been proposed include superconducting circuits and nano-mechanics \cite{11Sarlette,14NavarreteBenlloch,13Kronwald}. Reservoir engineering provides access to controlled dissipation which can be used in quantum simulations of open quantum systems \cite{14NavarreteBenlloch,00Myatt}.

\nocite{00Myatt}
\nocite{BkHaroche}
\nocite{00Fidio}
\nocite{76Yuen}
\nocite{13Alonso}
\nocite{00Turchette2}
\nocite{98Wineland2}
\nocite{00Sorensen1}

\bibliography{./myrefs}

\begin{thebibliography}{10}

\bibitem{93Cirac2}
J.~I. Cirac, A.~S. Parkins, R.~Blatt, P.~Zoller, {\it Phys. Rev. Lett.\/} {\bf
  70}, 556 (1993).

\bibitem{96Poyatos}
J.~F. Poyatos, J.~I. Cirac, P.~Zoller, {\it Phys. Rev. Lett.\/} {\bf 77}, 4728
  (1996).

\bibitem{01Carvalho}
A.~R.~R. Carvalho, P.~Milman, R.~L. de~Matos~Filho, L.~Davidovich, {\it Phys.
  Rev. Lett.\/} {\bf 86}, 4988 (2001).

\bibitem{08Verstraete}
F.~Verstraete, M.~M. Wolf, J.~I. Cirac, {\it Nature Physics\/} {\bf 5}, 633
  (2009).

\bibitem{08Diehl}
S.~Diehl, {\it et~al.\/}, {\it Nature Physics\/} {\bf 4}, 878 (2008).

\bibitem{08Kraus}
B.~Kraus, {\it et~al.\/}, {\it Phys. Rev. A\/} {\bf 78}, 042307 (2008).

\bibitem{11Pastawski}
F.~Pastawski, L.~Clemente, J.~I. Cirac, {\it Phys. Rev. A\/} {\bf 83}, 012304
  (2011).

\bibitem{11Krauter}
H.~Krauter, {\it et~al.\/}, {\it Physical Review Letters\/} {\bf 107}, 080503
  (2011).

\bibitem{11Barreiro}
J.~T. Barreiro, {\it et~al.\/}, {\it Nature\/} {\bf 470}, 486 (2011).

\bibitem{13Lin}
Y.~Lin, {\it et~al.\/}, {\it Nature\/} {\bf 504}, 415 (2013).

\bibitem{13Shankar}
S.~Shankar, {\it et~al.\/}, {\it Nature\/} {\bf 504}, 419 (2013).

\bibitem{11Sarlette}
A.~Sarlette, J.~M. Raimond, M.~Brune, P.~Rouchon, {\it Phys. Rev. Lett.\/} {\bf
  107}, 010402 (2011).

\bibitem{14NavarreteBenlloch}
C.~Navarrete-Benlloch, J.~J. Garc\'ia-Ripoll, D.~Porras, {\it Phys. Rev.
  Lett.\/} {\bf 113}, 193601 (2014).

\bibitem{13Kronwald}
A.~Kronwald, F.~Marquardt, A.~A. Clark, {\it Phys. Rev. A\/} {\bf 88} (2013).

\bibitem{BkSchleich}
W.~Schleich, {\it Quantum Optics in Phase Space\/} (Wiley, 2001).

\bibitem{Methods}
Materials and methods are available as supplementary materials on
  \emph{Science} {Online}.

\bibitem{ThRoos}
C.~Roos, Controlling the quantum state of trapped ions, Ph.{D}.~thesis,
  Universit\"{a}t Innsbruck, Austria (2000).

\bibitem{98Wineland2}
D.~J. Wineland, {\it et~al.\/}, {\it J. Res. Natl. Inst. Stand. Technol.\/}
  {\bf 103}, 259 (1998).

\bibitem{08RoosGate}
C.~F. Roos, {\it New J. Phys.\/} {\bf 10}, 013002 (2008).

\bibitem{96Meekhof}
D.~M. Meekhof, C.~Monroe, B.~E. King, W.~M. Itano, D.~J. Wineland, {\it Phys.
  Rev. Lett.\/} {\bf 76}, 1796 (1996). \emph{Phys. Rev. Lett.} \bf{77}\rm,
  2346(E) (1996).

\bibitem{76Yuen}
H.~P. Yuen, {\it Phys. Rev. A\/} {\bf 13}, 2226 (1976).

\bibitem{00Gottesman2}
D.~Gottesman, A.~Kitaev, J.~Preskill, {\it Phys. Rev. A\/} {\bf 64}, 012310
  (2001).

\bibitem{00Myatt}
C.~J. Myatt, {\it et~al.\/}, {\it Nature\/} {\bf 403}, 269 (2000).

\bibitem{BkHaroche}
S.~Haroche, J.-M. Raimond, {\it Exploring the Quantum: Atoms and Cavities and
  Photons\/} (Oxford University Press, (2006)).

\bibitem{00Fidio}
C.~Di~Fidio, W.~Vogel, {\it Phys. Rev. A\/} {\bf 62}, 031802 (2000).

\bibitem{13Alonso}
J.~Alonso, F.~M. Leupold, B.~C. Keitch, J.~P. Home, {\it New Journal of
  Physics\/} {\bf 15}, 023001 (2013).

\bibitem{00Turchette2}
Q.~A. Turchette, {\it et~al.\/}, {\it Phys. Rev. A\/} {\bf 62}, 053807 (2000).

\bibitem{00Sorensen1}
A.~S{\o}rensen, K.~M{\o}lmer, {\it Phys. Rev. A\/} {\bf 62}, 022311 (2000).

\end{thebibliography}

\bibliographystyle{Science}



\noindent {\bf Acknowledgements} We thank Joseba Alonso, Atac Imamoglu, and David Wineland for feedback on the manuscript and useful discussions. We thank Joseba Alonso, Martin Sepiol, Karin Fisher and Christa Fl{\"u}hmann for contributions to the experimental apparatus. We acknowledge support from the Swiss National Science Foundation under grant number $200021\_134776$, and through the National Centre of Competence in Research for Quantum Science and Technology (QSIT).\\
\\
\noindent {\bf Supplementary Materials} \\
Supplementary Text\\
Tables S1, S2, S3\\
References (\emph{24-28})\\



\clearpage
\begin{figure}
\includegraphics[angle=-90,width=\textwidth]{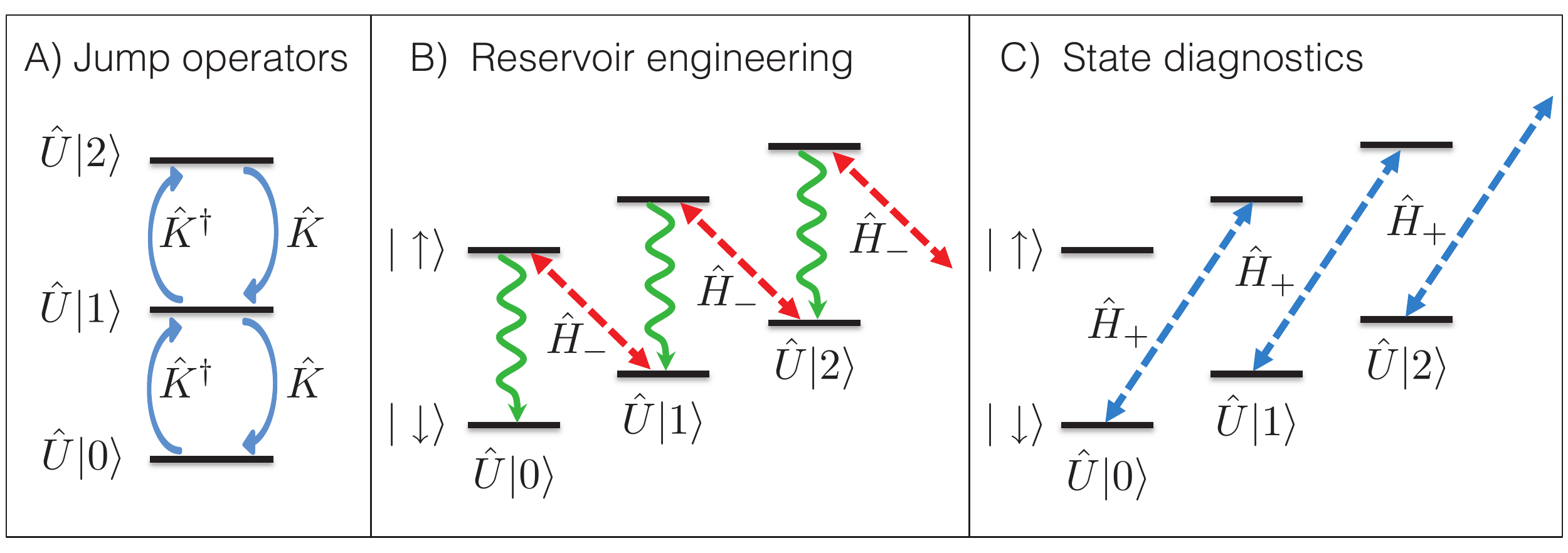}\\
\caption{{\bf Cooling and detection using an engineered state ladder.} A) The operators $\hat{K}$, $\hat{K}^\dagger$ are ladder operators for the set of basis states $\hat{U}\left|n\right>$.
B) A combination of the reversible Hamiltonian $\hat{H}_-$ and internal state dissipation leads to pumping into the ground state of the engineered basis. C) State measurement probes the $\hat{H}_+$ Hamiltonian, resulting in single frequency Rabi oscillations if the system is prepared in $\hat{U}\left|0\right>$}
\end{figure}

\begin{figure}
\includegraphics[width=\textwidth]{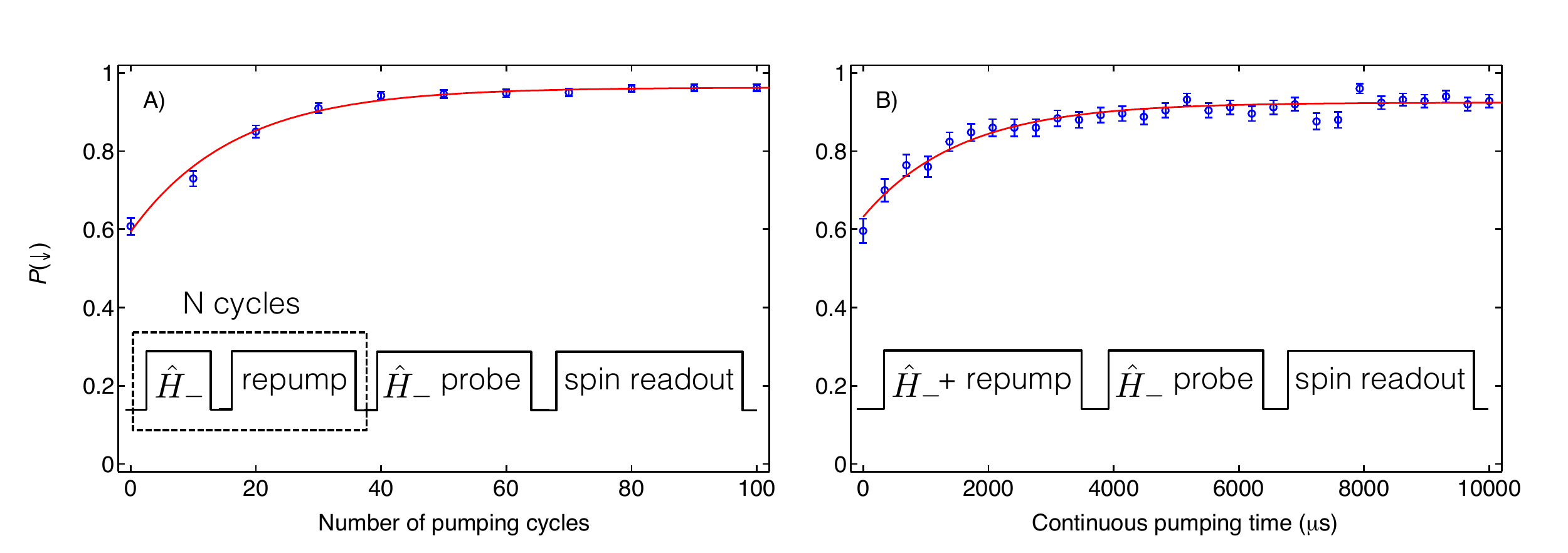}\\
\caption{{\bf Onset of the dark state of the reservoir engineering.} A) Pulsed preparation of a coherent state. Each pulse cycle consists of applying $\hat{H}_-$ for 25~$\mu$s and then repumping the spin. B) Continuous pumping into a squeezed state, in which $\hat{H}_-$ is turned on simultaneously with optical pumping of the spin.  The $\hat{H}_-$ analysis pulse lengths are 80~$\mu$s and 55~$\mu$s respectively. In all figures the data points are the mean measured spin population based on $> 300$ repetitions of the experimental sequence, with error bars estimated from quantum projection noise.}
\end{figure}

\begin{figure}
\includegraphics[width=\textwidth]{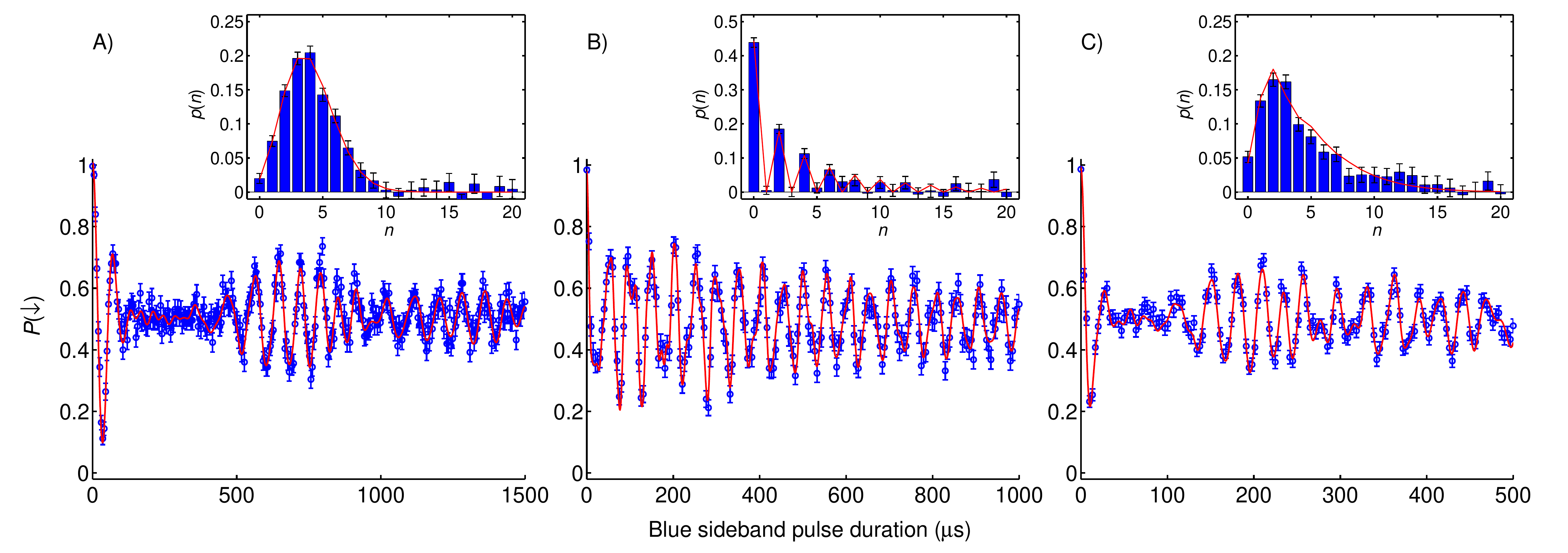}\\
\caption{{\bf Fock state analysis using a single-frequency blue-sideband probe.} All data are fitted using a form similar to equation \ref{eq:Pflop} to retrieve the Fock state populations shown in the inset bar charts. The red curves in the bar charts are fits to the populations using the relevant probability distribution $p(n)$ to determine the size of the states. Data, populations and fitted distributions are shown for  A) the coherent state (with fitted $\left|\alpha\right| = 2.00\pm0.01$), 
 B) the squeezed vacuum state ($r = 1.45\pm0.03$) and  
 C) the displaced-squeezed state ($\left|\alpha\right| = 2.2\pm0.2$, $r = 0.63\pm0.06$ and $\arg(\alpha) - \phi_{\rm s}/2 = 0.42\pm0.06$~rad).} 
\end{figure}

\begin{figure}
\includegraphics[width=\textwidth]{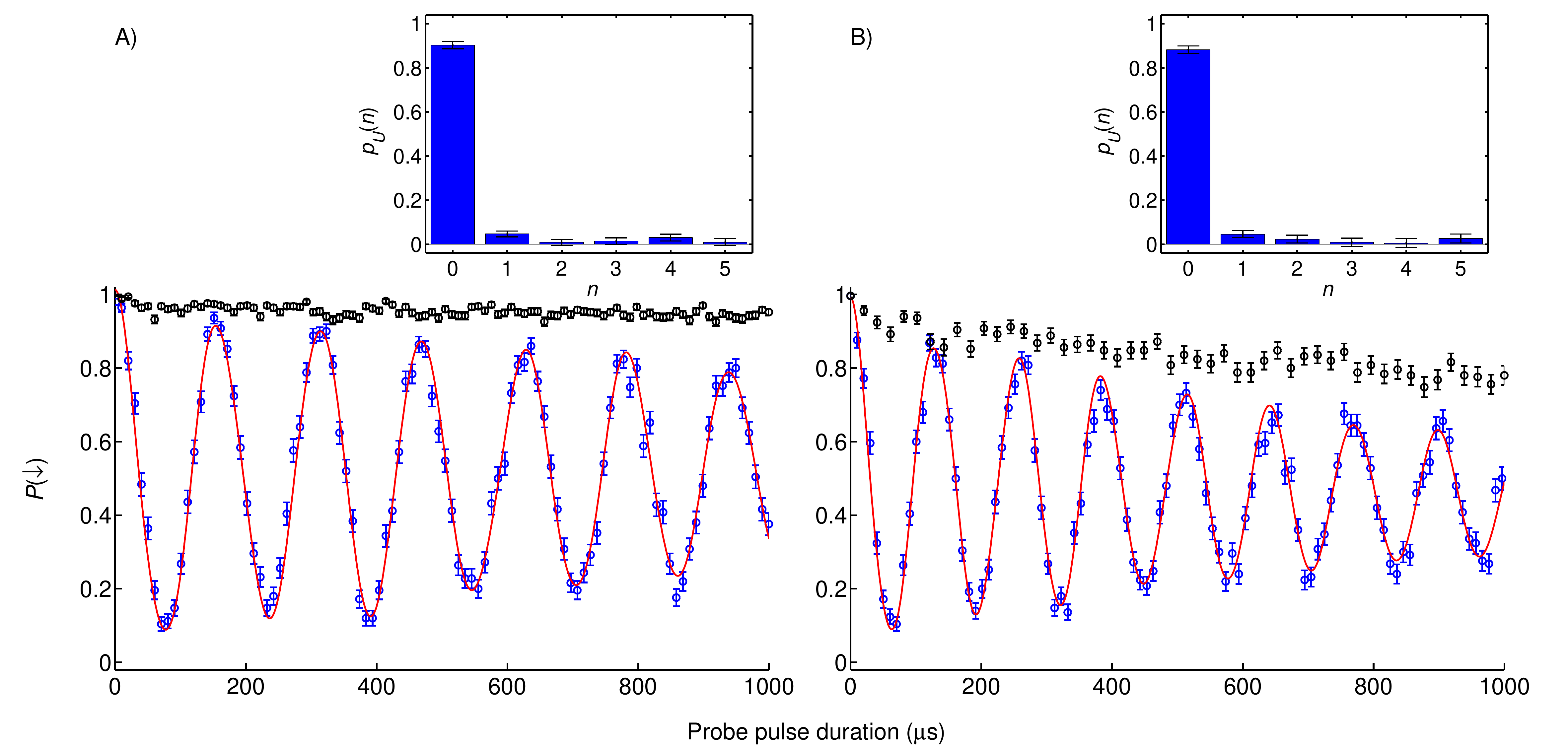}\\
\caption{{\bf Fock state analysis in the engineered basis.} The data show the spin population evolution after applying $\hat{H}_+$ (blue data points) and $\hat{H}_-$ (black data points) for time $t$. The fitted curves for $\hat{H}_+$ are of a form similar to equation \ref{eq:Pflop} \cite{Methods}. The resulting populations are shown in the inset bar charts. Data and populations are shown for A) the coherent state, with fitted $p_U(0) = 0.90\pm0.02$, and 
    B) the squeezed vacuum state, with $p_U(0) = 0.88\pm0.02$.} 


\end{figure}

\clearpage
\section{Supplementary Information}

\subsection{Introduction}

In this Supplementary Information we give technical information, analyze in depth the limitations of the experiment, and describe the data analysis. A description of the experimental setup is given in section \ref{sec:Experimental setup} and the experimental implementation of the $\hat{H}_-$ and $\hat{H}_+$ Hamiltonians in section \ref{sec:Laser frequency comp}. Section \ref{sec:Dissipative pumping rate} discusses the pumping rate into the ground state. A description of the data analysis methods are given in section \ref{sec:Fits}. We provide a discussion of limits to the squeezed state size due to sources of noise in our experiment in section \ref{sec:Heating and dephasing}. The validity of the Lamb-Dicke approximation and its implications for the ultimate size of states which might be achievable in our setup using reservoir engineering is discussed in section \ref{sec:LD approx}. For comparison we deduce the limits for the state fidelity of a squeezed state created with coherent rather than dissipative means in section \ref{sec:Coherent generation sqs}.

The mathematical definitions in the main text differ from those used in the Supplementary Information. Table S1 summarizes the definitions used.

\begin{table}[h]
    \caption*{Table S1: Summarized notation differences between main text and Supplementary Information (S.I.).}\label{table:defs}
    \centering
    \begin{tabular}{| l | l | l |}
    \hline
    Name & Definition paper & Definition S.I. \\ \hline
    Carrier Hamiltonian & $\hat{H}_{\rm c} = \hbar \Omega_{\rm c} \splus + {\rm h.c.}$ & $\hat{H}_{\rm c} = \hbar \Omega_{\rm c} \splus e^{i \phi_{\rm c}} + {\rm h.c.}$ \\ \hline
    Rabi frequency (carrier)& $|\Omega_{\rm c}|$ & $\Omega_{\rm c}$ \\ \hline
    Phase (carrier) & $\arg(\Omega_{\rm c})$ & $\phi_{\rm c}$ \\ \hline
    Red sideband Hamiltonian & $\hat{H}_{\rm rsb} = \hbar \Omega_{\rm rsb} \splus \destroy +  {\rm h.c.}$ & $\hat{H}_{\rm rsb} = \hbar \Omega_{\rm rsb} \splus \destroy e^{i \phi_{\rm rsb}} +  {\rm h.c.}$\\ \hline
    Rabi frequency (red sideband)& $|\Omega_{\rm rsb}|$ & $\Omega_{\rm rsb}$ \\ \hline
    Phase (red sideband) & $\arg(\Omega_{\rm rsb})$ & $\phi_{\rm rsb}$ \\ \hline
    Blue sideband Hamiltonian & $\hat{H}_{\rm bsb} = \hbar \Omega_{\rm bsb} \splus \create +  {\rm h.c.}$ & $\hat{H}_{\rm bsb} = \hbar \Omega_{\rm bsb} \splus \create e^{i \phi_{\rm bsb}} +  {\rm h.c.}$\\ \hline
    Rabi frequency (blue sideband)& $|\Omega_{\rm bsb}|$ & $\Omega_{\rm bsb}$ \\ \hline
    Phase (blue sideband) & $\arg(\Omega_{\rm bsb})$ & $\phi_{\rm bsb}$ \\ \hline
    \end{tabular}
\end{table}

\subsection{Experimental setup} \label{sec:Experimental setup}
The laser beams addressing the dipole transitions of the $^{40}\rm{Ca}^+$ ion which are used for Doppler cooling, internal state detection and optical pumping have the wavelengths 397, 854 and 866~nm.
The laser beam used for manipulating the pseudo-spin addresses the quadrupole transition $\ket{S_{1/2}, M_J = +1/2} \leftrightarrow \ket{D_{5/2}, M_J = +3/2}$ at 729~nm. It is incident on the ion at an angle of 45 degrees to the ion's axial motion (Lamb-Dicke parameter $\eta = 0.05$) and parallel to the applied magnetic field of 11.9~mT. In this high field Doppler cooling requires multiple frequency components for both the 866 and 397~nm lasers. We measure the mean energy eigenstate occupation after Doppler cooling to be $\bar{n} \approx 10$.

Optical pumping of the internal state is performed by frequency selectivity using a 397~nm beam containing equal amounts of $\sigma^+$ and $\sigma^-$ polarization components (this beam is directed perpendicular to the magnetic field). The $\sigma^+$ polarization component of the 397~nm light is resonant with the $\ket{S_{1/2}, M_J = -1/2} \leftrightarrow \ket{P_{1/2}, M_J = +1/2}$ transition. Since the $S_{1/2}$ states are split by $334~$MHz and the $P_{1/2}$ states are split by $110$~MHz in the field of 11.9~mT, the $\sigma^-$ polarization component is 444~MHz from resonance. At one saturation intensity, the state preparation fidelity can be calculated to be 0.997. At 1/40 of a saturation intensity, rate equation simulations indicate that fidelities of 0.9993 are achievable.

We read out the internal state of the ion by observing the level of fluorescence when applying resonant laser fields at 397 and 866~nm. The number of detected 397~nm photons follows a Poisson distribution which for an initial $\ket{\downarrow} (\ket{\uparrow})$ state has a mean of 30 counts (0.5 counts) in a detection time of 300~$\mu$s.

\subsection{Laser frequency components for $\hat{H}_-$ and $\hat{H}_+$}\label{sec:Laser frequency comp}
The coherent (Glauber-Sudarshan) state $\ket{D(\alpha),0}$ is defined by the application of the displacement operator $\hat{D}(\alpha) \equiv e^{\alpha \create - \alpha^* \destroy}$ to the ground state $\ket{0}$. In order to create a coherent state by reservoir engineering, we implement $\hat{H}_-$ with $\hat{K} = \hat{D}(\alpha) \destroy \hat{D}(-\alpha) = \destroy - \alpha$. This involves simultaneously applying two laser fields, one which resonantly drives the carrier transition $\ket{\downarrow} \ket{n} \leftrightarrow \ket{\uparrow}\ket{n}$ and another which resonantly drives the red motional sideband transition $\ket{\downarrow} \ket{n} \leftrightarrow \ket{\uparrow}\ket{n - 1}$. This realizes the Hamiltonian $\hat{H}_-$ with $|\alpha| = \Omega_{\rm c}/\Omega_{\rm rsb}$ and $\arg(\alpha) = \phi_{\rm c} - \phi_{\rm rsb} + \pi$ where $\Omega_{\rm c}$, $\phi_{\rm c}$ and $\Omega_{\rm rsb}$, $\phi_{\rm rsb}$ are the Rabi frequencies and phases for the carrier and red sideband transition respectively. We realize $\hat{H}_+$ by applying a bichromatic optical field which has frequency components resonant with both the carrier transition and the blue sideband, with the ratio of Rabi frequencies $|\Omega_{\rm c}/\Omega_{\rm bsb}| = |\alpha|$ and the relative phase $\phi_{\rm c} - \phi_{\rm bsb} + \pi = -\arg(\alpha)$.

In experiments working only with coherent states, the multiple frequency components were generated by splitting the light into two paths and passing it through two double-pass acousto-optic modulators. These two paths were then combined on a beamsplitter before coupling into a single mode fibre which delivers the light to the trap.

The squeezed vacuum state $\ket{\hat{S}(\xi),0}$ is defined by the application of the squeezing operator $\hat{S}(\xi) \equiv e^{( \xi^* {\destroy}^2 -  \xi {\create}^2)/2}$ to the ground state $\ket{0}$. The complex number $\xi$ can be written in terms of its magnitude $r$ and phase $\phi_{\rm s}$. In order to generate squeezed vacuum states as the dark state of a pumping process, we implement a coupling Hamiltonian of the form given in equation 1 in the main text, with the operator $\hat{K} = \hat{S}(\xi) \destroy \hat{S}^\dagger(\xi) = \cosh(r) \destroy + e^{i\phi_{\rm s}} \sinh(r) \create$. This requires simultaneous application of laser fields resonant with both the red motional sideband transition and the blue motional sideband transition $\ket{\downarrow} \ket{n} \leftrightarrow \ket{\uparrow}\ket{n + 1}$. The ratio of the Rabi frequencies on these transitions gives the strength of the squeezing through $\tanh(r) = |\Omega_{\rm bsb}/\Omega_{\rm rsb}|$ while the phase is given by the phase difference between the two frequency components of the laser $\phi_{\rm s} = \phi_{\rm bsb} - \phi_{\rm rsb}$.

For the squeezed state, we implement $\hat{H}_+$ by simultaneously applying laser fields resonant with the blue and red motional sidebands with a ratio of Rabi frequencies $|\Omega_{\rm rsb}/\Omega_{\rm bsb}| = \tanh(r)$, and with a phase difference $\phi_{\rm bsb} - \phi_{\rm rsb} = \phi_{\rm s}$.

The displaced-squeezed states can be defined as a combination of a displacement and a squeezing operator acting on the ground state, $\hat{S}(\xi) \hat{D}(\alpha) \ket{0}$. To create displaced-squeezed states we add a component resonant with the carrier transition with Rabi frequency $\Omega_{c}$ to the squeezed state Hamiltonian, creating the operator $\hat{K} = \cosh(r) \destroy + e^{i\phi_{\rm s}} \sinh(r) \create - \alpha$ with $r$ and $\phi_{\rm s}$ defined as for the squeezed state, and $\Omega_{\rm c} e^{i\phi_{\rm c} - i\phi_{\rm rsb}}/\Omega_{\rm rsb} = -\alpha/\cosh(r)$.

For the squeezed states and the displaced-squeezed states we applied the three frequency components as radio-frequency tones to a single-pass acousto-optic modulator, after which all components of the light were coupled into the same optical fibre. In order to achieve good coupling of all three, the centre of the acousto-optic modulator was imaged onto the fibre.

\subsection{Dissipative pumping rate: continuous pumping}\label{sec:Dissipative pumping rate}
The result of applying $\hat{H}_-$ is to induce Rabi oscillations with frequency $\Omega$ between the spin states which are correlated with transitions in the engineered basis $\ket{\downarrow} \ket{\hat{U}, n} \leftrightarrow \ket{\uparrow} \ket{\hat{U}, n - 1}$. The optical pumping relaxes the spin from $\ket{\uparrow}$ to $\ket{\downarrow}$ at rate $\Gamma$ (figure 1
b). In the Lamb-Dicke regime the atomic recoil associated with optical pumping can be neglected, and the spin system forms a zero temperature reservoir for the motion \cite{00Myatt}. If both the Hamiltonian and optical pumping are applied simultaneously with $\Gamma \gg \Omega$ \cite{BkHaroche}, the excited spin state can be adiabatically eliminated, and the resulting evolution of the motional state density operator $\hat{\rho}_m$ is described by a Master equation in the Lindblad form
\be
\frac{d \hat{\rho}_m}{dt} = \Gamma_m \left[\hat{K} \hat{\rho}_m \hat{K}^\dagger - \frac{1}{2} \left(\hat{K}^\dagger \hat{K} \hat{\rho}_m + \hat{\rho}_m \hat{K}^\dagger \hat{K} \right)\right] \label{eq:Master},
\ee
where $\Gamma_m = 2 \Omega^2/\Gamma$.

For the squeezed state, the dissipative pumping slows down as the size of squeezed state which we try to engineer is increased. The relation between the blue and red sideband Rabi frequencies and the Rabi frequency which is relevant to the pumping rate $2 \Omega^2/\Gamma$ for the squeezed state can be seen from
\be
\hat{H}_{\rm sq} &=& \hbar \Omega \cosh(r)\left( \create + \tanh(r) e^{i\phi_s} \destroy \right) \hat{\sigma}_+ + {\rm h.c.} \nonumber \\ &=& \hbar\left(\Omega_{\rm bsb} e^{i \phi_{\rm bsb}} \create + \Omega_{\rm rsb} e^{i \phi_{\rm rsb} } \destroy\right)\hat{\sigma}_+ + {\rm h.c.} \ .
 \ee
The pumping rate scales as $\Omega_{\rm bsb}^2/\cosh^2(r)$.

\subsection{Fits to Rabi oscillation data}\label{sec:Fits}
To extract the populations $p(n)$ of the Fock states fits to data were performed using a fitting function of the form
\be
P(\downarrow,t) =  b t + \frac{1}{2}\sum_{n = 0}^{n_{\rm max}} p(n)\; \left(1+ \ e^{-\gamma \sqrt{n+1} t} \cos(\Omega_{n, n + 1} t)\right).
\label{eq:Pflopfit}
\ee
where $\Omega_{n,n+ 1} = \Omega_{\rm R} f(n, \eta)$ is the state-dependent Rabi frequency. $\Omega_{\rm R}$ is a constant proportional to the square root of the laser intensity. We use a scaling function $f(n, \eta) = e^{-\eta^2/2}(1/(n + 1))^{1/2} \eta L_n^{1}(\eta^2)$ which is known to be correct when driving the blue sideband. At low $n$, $f(n, \eta) \propto \sqrt{n + 1}$, both for the blue sideband and for the $\hat{H}_+$ Hamiltonian. For higher values of $n$ we expect that the explicit form of $f(n, \eta)$ would be different for the blue sideband and $\hat{H}_+$. In the latter case, populations with $n>0$ are not observed in our experiments.

The parameter $b$ in the first term on the right hand side of equation \ref{eq:Pflopfit} accounts for a gradual pumping of population into the state which is not involved in the dynamics of the probe Hamiltonian \cite{00Fidio}. This occurs due to frequency noise components in our laser which are resonant with the carrier transition (due to finite servo bandwidth). This pumping predominantly occurs as a direct spin flip with no effect on the motion. Thus the effect is largest when the population of the motional ground state of the basis relevant to the probe Hamiltonian is large. This is most significant for the data shown in figure 4, where it gives a $~5$\% effect over the 1~ms duration of the Rabi oscillations. For the data shown in figure 3, this is a smaller effect, and we fix $b$ to zero for fitting.

Once we have extracted $p(n)$ from the data, we fit these populations with the theoretical probability distributions obtained from the general form for a displaced-squeezed state \cite{76Yuen}
\be
p_{\rm ds}\left(n, r, \phi_{\rm s}, \alpha \right) = \frac{(\tanh(r)/2)^n}{n! \cosh(r)} e^{-|\alpha|^2 + |\alpha|^2 \tanh(r) \cos(2 \arg(\alpha) - \phi_{\rm s})} \left| H_{n}\left(\frac{|\alpha|  e^{i (\arg(\alpha) - \phi_{\rm s}/2)}}{\sqrt{\sinh(2 r)}} \right) \right|^2
\label{eq:Pyuen}
\ee
where $H_n(x)$ is a Hermite polynomial, $\alpha = \left|\alpha\right| e^{i \arg(\alpha)}$ and $\xi = r e^{i \phi_{\rm s}}$.  For the coherent state generation we take the limit of this expression as $r\rightarrow 0$, resulting in the  Poisson distribution
\be
p_{\rm c}(n, |\alpha|) = \frac{e^{-|\alpha|^2}|\alpha|^{2n}}{n!}\ .
\ee
For the squeezed vacuum state we take $|\alpha| = 0$, giving the probability distribution
\be
p_{\rm s}(n, r) &=& \frac{(\tanh(r)/2)^n n!}{((n/2)!)^2 \cosh(r)},\ n~\textrm{ even}\\
p_{\rm s}(n, r) &=& 0,\ n~\textrm{ odd.}
\ee

\begin{landscape}
\subsubsection{Fitting results}
\begin{table}[h]
    \caption*{Table S2: Combined fitting results for data using a single-frequency probe on the blue sideband transition. Fit parameters indicated by ``-'' were fixed to zero for the fit to the population.
    }
    \centering
    \begin{tabular}{| l | l | l | l | l | l | l | l | l | l | l |}
    \hline
    Data set                   & $b$ (1/s)& $|\alpha|$   & $r$         & $\arg(\alpha) - \phi_{\rm s}/2$& $\gamma$ (1/ms) & $\Omega_{\rm R}/(2\pi)$ (kHz)&$n_{\rm max}$\\ \hline
    Coherent state             & - & $2.00\pm0.01$    & -         & -                             & $0.37\pm0.02$ & $128.50\pm0.06$       &20\\ \hline
    Squeezed vacuum state      & - & -          & $1.45\pm0.03$   & -                             & $0.99\pm0.07$ & $404.6\pm0.2$       &20\\ \hline
    Displaced squeezed state   & - & $2.2\pm0.2$     & $0.63\pm0.06$   & $0.42\pm0.06$                       & $0.74\pm0.08$  & $395.8\pm0.2$       &20\\ \hline
    \end{tabular}
\end{table}

\begin{table}[h]
    \caption*{Table S3: Fitting results for data using $\hat{H}_+$ as probe pulse. Equation \ref{eq:Pflopfit} was used for the fitting. 
    }
    \centering
    \begin{tabular}{| l | l | l | l | l | l | l | l |}
    \hline
    Data set               & $b$ (1/s) & $p_U(0)$     & $\gamma$ (1/ms)  & $\Omega_{\rm R}/(2 \pi)$ (kHz) &$n_{\rm max}$\\   \hline
    Coherent state         & $19\pm9$  & $0.90\pm2$   & $0.52\pm0.04$& $130.3\pm0.1$&5\\   \hline
    Squeezed vacuum state  & $-49\pm4$ & $0.88\pm2$   & $1.05\pm0.04$& $159.0\pm0.1$&5\\   \hline
    \end{tabular}

\end{table}

\end{landscape}

\section{Heating and dephasing of squeezed vacuum states}\label{sec:Heating and dephasing}

The rate at which a squeezed state loses overlap with itself due to fluctuating electric fields at the ion is related to the heating rate from the ground state $\Gamma_{0 \rightarrow 1}$ by
$
\Gamma_{\rm sq} = \Gamma_{0\rightarrow 1} \frac{\cosh(2 r)}{2} \
$ \cite{13Alonso}.
This form invokes the assumption that the noise is well described by a stationary random variable with a short correlation time. For our observed heating rate of the axial motional mode from its quantum ground state of $\Gamma_{0 \rightarrow 1} = 10\pm1$~quanta~s$^{-1}$ this results in a rate of $46$~quanta~s$^{-1}$ for a squeezed state with $r = 1.45$. This is slower than the timescale of decoherence observed in both the $\hat{H}_-$ and $\hat{H}_+$ data shown in figure 4
. We have performed Monte-Carlo wavefunction simulations using Lindblad operators $\sqrt{\Gamma_{0 \rightarrow 1}} \destroy$ and $\sqrt{\Gamma_{0 \rightarrow 1}} \create$ to verify this observation.

In our current trap, the motional coherence time of the Fock state superposition $\left(\ket{0} + \ket{1}\right)/\sqrt{2}$ has been measured a month before the squeezed state data to be $32\pm3$~ms, with an approximate exponential form. This may vary over the timescales of months over which the data was taken. This measurement is not consistent with the measured heating rate, since the coherence time scales as $e^{-2 \Gamma_{0 \rightarrow 1 }t}$ due to amplitude damping \cite{00Turchette2}. This indicates that there may be an as yet unidentified source of motional dephasing in our trap. Motional dephasing at a level which is hard to observe for the Fock state superposition could have a large affect on squeezed state decoherence. We can estimate the decay parameter due to the dephasing using the difference between the observed motional coherence and heating rates $\Gamma_{\rm dephase} = 1/(32~{\rm ms}) - 2\Gamma_{0 \rightarrow 1}$. When we use this value in Monte-Carlo wavefunction simulations of the $\hat{H}_+$ and $\hat{H}_-$ evolutions with an operator $\Gamma_{\rm dephase} \create \destroy$ the loss of contrast is faster than what we observe experimentally for the squeezed state. We thus conclude that dephasing is likely to be a major factor in the decay of oscillations observed in the data in figure 4
. The assumption of Markovian dephasing may not correctly describe our experiment; the real scaling between decoherence rates for the two-state superposition and the squeezed state will depend on the characteristics of the noise which is causing this effect.

Though dephasing limits the time over which we observe Rabi oscillations, Monte-Carlo wavefunction simulations of the pumping including the effects of dephasing, heating, and the Rabi frequencies used in the experiments indicate that these mechanisms should not limit the size of the squeezed vacuum state which we can produce at the current levels. We have tried to produce squeezed states with $r$ up to 2, but we cannot currently reconcile the data for these with simulations.

\subsection{Validity and limits of the Lamb-Dicke approximation}\label{sec:LD approx}
The validity of the Lamb-Dicke approximation presents a fundamental limit on the size of squeezed states which can be reached using reservoir engineering. We are able to estimate the largest state which can be produced with high fidelity by looking at the dark state of the resonant terms in the actual Hamiltonian which we produce in the laboratory, which is given by $
\hat{H}_{\rm lab} = \left( \Omega_{n, n + 1} \sinh(r) \ket{n}\bra{n + 1} + \Omega_{n, n - 1} \cosh(r) \ket{n}\bra{n - 1} + {\rm h.c.}\right)
$
with $\Omega_{n, n + 1} = \Omega_{0,0} (1/(n + 1))^{1/2} \eta L_n^{1}(\eta^2)$ where $L_n^\alpha(x)$ is the generalized Laguerre polynomial in $x$ \cite{98Wineland2}. For the Lamb-Dicke parameter used in our experiments $(\eta = 0.05)$ the dark state fidelity with the desired squeezed state drops below $0.95$ for $r = 2.9$, which is well above the regime accessed in our experiments. Population starts to accumulate in the vicinity of the Fock state for which $\Omega_{n, n+1}$ is close to or equal to zero -- this means that although the theoretical fidelity is 0.95 there is already a significant increase in the variance of the squeezed quadrature compared to what would be expected for an ideal squeezed vacuum state. The maximum reduction in the squeezed variance achievable with this value of the Lamb-Dicke parameter is 21~dB. It is interesting to note that the squeezed state being dark is due to the ratio of the neighbouring sideband matrix elements $\Omega_{n, n+1}/\Omega_{n, n-1}$. Though this is strictly valid only in the Lamb-Dicke regime, it is also well approximated for significant deviations from the Lamb-Dicke $\sqrt{n}$ scaling of the matrix elements, meaning that the state still closely approximates the desired squeezed state.

The Lamb-Dicke approximation is also implicit in the assumption that the photons scattered during the internal state repumping do not induce significant changes to the motional state. This reduces the pumping rate, since it adds a diffusion process, but it does not effect the dark state in the absence of other heating mechanisms (the scattering stops once the system attains the dark state). In practice, off-resonant driving of the carrier transition results in residual internal state excitation even if the motion is in the nominal dark state. This means that the recoil due to scattering can degrade the motional steady-state fidelity. For the scattering, the relevant Lamb-Dicke parameter is that for the scattered photons, which have wavelengths of $393$~nm, $397$~nm and $854$~nm. The higher momentum of the ultra-violet photons would be expected to introduce more problems, and an average of 3 ultraviolet photons are scattered during repumping for our setup. To gauge the influence which both effects have, we have performed Monte-Carlo wavefunction simulations which include off-resonant driving of the carrier transition, and in which the momentum kick during relaxation is included using a Lindblad operator $ \hat{\sigma}_- e^{\chi i \eta_{\rm 393} (\create + \destroy)}$. Values of $\chi > 1$ were used to account for the multiple scattering events. We obtain fidelities of $0.95\pm1$ for a squeezed state with $r = 1.38$ in the deliberately chosen extreme case where $\chi = 3$, which is similar to assuming that all photons involved in the repumping pulse displace the ion in the same direction, and where this is taken to be along the oscillator axis. This most likely overestimates the effect of the recoil. Nevertheless, this fidelity is higher than that we observe in the experiment.

\subsection{Coherent generation of squeezed states}\label{sec:Coherent generation sqs}
An alternative approach to generating squeezed states would be to prepare the ground state, and then subsequently to apply a suitable Hamiltonian for a fixed duration. The desired Hamiltonian is
\be
\hat{H} = \frac{\hbar \Omega}{2} \left({\create}^2 + \destroy^2 \right) \ ,
\ee
which should act only on the motional state. This will produce a squeezed state with squeezing amplitude $r$ in a time of $t_r = r/\Omega$. In order to generate this Hamiltonian using our 729~nm laser, one possibility is to prepare an eigenstate of the spin $\sigma_x$ operator, then drive both second motional sidebands simultaneously. This is a modification of the methods used for generating the displacement operators used in the two-qubit gate proposed by S\o rensen and M\o lmer \cite{00Sorensen1}. The strength of the second sideband is given relative to that of the first sideband by $\eta$. For our experiment, typical values of first sideband Rabi frequencies are $\Omega_{\rm bsb}/(2 \pi) = 20$~kHz, hence the Rabi frequency for the second sideband would be $\Omega/(2 \pi) = 1$~kHz. Generating a squeezed state with $r = 1.45$ would therefore take $230~\mu$s. The $\hat{H}_+$ oscillations displayed in figure 4 (b) are at less than $80\%$ of their initial amplitude over this timescale, which is lower than the fidelity observed in our experiments.



\end{document}